\documentclass[%
 reprint,
 amsmath,amssymb,
 aps,
prb,
 floats,
]{revtex4-1}

\usepackage{subfigure}
\usepackage{graphicx}
\graphicspath{{./}{./images/}{./images_all/}{./images_all/metastable/}}
\DeclareGraphicsExtensions{.eps}

\usepackage{dcolumn}
\usepackage{bm}

\begin{document}

\preprint{APS/123-QED}

\title{Metastable state
in a shape-anisotropic single-domain nanomagnet \\ subjected to spin-transfer-torque}


\author{Kuntal Roy$^{1}$}
\email{royk@vcu.edu}
\author{Supriyo Bandyopadhyay$^1$}
\author{Jayasimha Atulasimha$^2$}
\affiliation{$^1$Dept. of Electrical and Computer Engr., Virginia Commonwealth University, Richmond, VA 23284, USA\\
$^2$Dept. of Mechanical and Nuclear Engr., Virginia Commonwealth University, Richmond, VA 23284, USA}

\date{\today}

\begin{abstract}
We predict the existence of a new metastable  magnetization state in a single-domain nanomagnet with uniaxial shape anisotropy.
It emerges when a spin-polarized 
current, delivering a spin-transfer-torque, is injected into the nanomagnet.  It can trap the magnetization vector and prevent  
spin-transfer-torque from switching the magnetization from one stable 
state along the easy axis to the other. Above a certain threshold current, the metastable state no longer appears.
This has important
technological consequences for spin-transfer-torque based magnetic memory and logic systems. 
  \end{abstract}


\maketitle


Spin-transfer-torque (STT) is an electric 
current-induced magnetization switching mechanism that is widely used to switch the 
magnetization of a  nanomagnet with uniaxial shape anisotropy
 from one stable state to the other~\cite{RefWorks:8,RefWorks:7}.  A spin-polarized current is injected
into the magnet to
deliver a torque on the magnetization vector and make it switch. 
This has now become the staple of
nonvolatile magnetic random access memory (STT-RAM) technology~\cite{RefWorks:435}.

In this Communication, we show analytically that the spin polarized current can 
spawn a metastable state in the magnet, which can trap the magnetization vector 
and prevent it from switching. This happens only if the spin-polarized current 
is smaller than a certain value. Thus, a minimum current -- which may be larger than the 
critical switching current -- may be needed for fail-safe switching.

\begin{figure}
\includegraphics[width=2.2in]{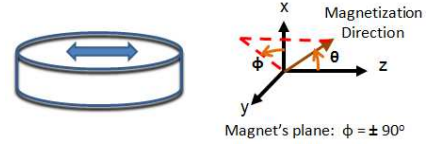}
\caption{\label{fig:single_magnet} A nanomagnet shaped like an elliptical cylinder. 
The nanomagnet cross-sections are on the y-z plane. 
The magnetization direction  can be rotated with a spin polarized current. }
\end{figure}

Consider a single-domain nanomagnet shaped like an elliptical cylinder with elliptical 
cross section in the y-z plane (see Fig.~\ref{fig:single_magnet}). The major (easy) and the minor (in-plane hard)
axes of the ellipse are aligned along the z-direction and y-direction, respectively. 
 Let $\theta(t)$ be the polar angle and $\phi(t)$  the azimuthal angle of the magnetization vector
in spherical coordinates. 

At any instant of time $t$, the energy of the unperturbed nanomagnet is the uniaxial shape anisotropy energy
which can be expressed as \cite{roy11_4_2}:
\begin{equation}
E(t) = B(t) sin^2\theta(t) + constant\;term,
\end{equation}
\noindent
where 
\begin{eqnarray}
B(t) &=& B(\phi(t)) = \frac{\mu_0}{2} \, M_s^2 \Omega \left\lbrack N_{d-xx} cos^2\phi(t) \right. \nonumber\\
&& \qquad\qquad \left. + N_{d-yy} sin^2\phi(t) - N_{d-zz}\right\rbrack. \label{eq:B_phi_t}
\end{eqnarray}
\noindent
Here $M_s$ is the saturation magnetization, $N_{d-xx}$, $N_{d-yy}$ and
$N_{d-zz}$ are the x-, y- and z-components of demagnetization factor~\cite{RefWorks:157},
and $\Omega$ is the nanomagnet's volume.  

The magnetization \textbf{M}(t) of the single-domain nanomagnet has a constant magnitude
but a variable orientation, so that we can represent it by the vector of unit norm $\mathbf{n_m}(t) 
=\mathbf{M}(t)/|\mathbf{M}| = \mathbf{\hat{e}_r}$ where $\mathbf{\hat{e}_r}$ is the unit vector in 
the radial direction. 
The other two unit vectors are denoted by 
$\mathbf{\hat{e}_\theta}$ and $\mathbf{\hat{e}_\phi}$ for $\theta$ and $\phi$ rotations, respectively.

The torque acting on the magnetization within unit volume due to shape anisotropy is \cite{roy11_4_2}
\begin{eqnarray}
\mathbf{T_E} (t) 
								 &=& - \mathbf{\hat{e}_r} \times \nabla E(\theta(t),\phi(t))   \nonumber\\
								 &=& - \{2 B(t) sin\theta(t) cos\theta(t)\} \mathbf{\hat{e}_\phi} - \{B_{0e}(t) \, sin\theta (t)\} \mathbf{\hat{e}_\theta}, \nonumber\\
\label{eq:torque_gradient}
\end{eqnarray}
\noindent
where
\begin{equation}
B_{0e}(t)=B_{0e}(\phi(t))=\frac{\mu_0}{2} \, M_s^2 \Omega (N_{d-xx}-N_{d-yy}) sin(2\phi(t)).
\label{eq:B_0e}
\end{equation}

Passage of a constant spin-polarized current $I$ perpendicular to the plane of the nanomagnet generates a
 spin-transfer-torque that is given 
by~\cite{roy11_4_2}
\begin{equation}
\mathbf{T_{STT}}(t) = s \, \left\lbrack c_{s}(V) \, sin\theta(t) \, \mathbf{\hat{e}_\theta} - b_{s}(V) \, sin\theta(t) \, \mathbf{\hat{e}_\phi} \right\rbrack,
\label{eq:torque_STT}
\end{equation}
\noindent	
where $s = (\hbar/2e)\eta I$ is the spin angular momentum deposition per unit time and $\eta$ 
is the degree of spin-polarization in the current $I$. The coefficients $b_{s}(V)$ and $c_{s}(V)$ are 
voltage-dependent dimensionless terms that arise 
when the nanomagnet is coupled with an insulating layer as in an MTJ~\cite{RefWorks:112} and the spin-polarized current tunnels
through this layer. We will use constant values of $b_{s}(V)$ and $c_{s}(V)$ for simplicity~\cite{RefWorks:403}. Furthermore, 
we will assume $b_{s}(V)=0.3\,|c_{s}(V)|$ and $|c_{s}(V)|=1$ to be in approximate agreement with the experimental results 
presented in Refs.~[\onlinecite{RefWorks:14}, \onlinecite{RefWorks:15}]. For $\theta$ = 180$^\circ$ to 0$^\circ$ switching, $c_{s}(V)=+1$, and for $\theta$ = 0$^\circ$ to 180$^\circ$ switching, $c_{s}(V)=-1$, while $b_{s}(V)=+0.3$ for both cases.

The magnetization dynamics of the single-domain nanomagnet under the action of various torques is 
described by the Landau-Lifshitz-Gilbert (LLG) equation as
\begin{multline}
\frac{d\mathbf{n_m}(t)}{dt} - \alpha \left(\mathbf{n_m}(t) \times \frac{d\mathbf{n_m}(t)}
{dt} \right) = -\frac{|\gamma|}{M_V}\,\mathbf{T_{eff}}(t)
\label{LLG}
\end{multline}
\noindent
where $\mathbf{T_{eff}}(t) = \mathbf{T_E}(t) + \mathbf{T_{STT}}(t)$, $\alpha$ is the dimensionless phenomenological Gilbert damping constant, 
$\gamma = 2\mu_B \mu_0/\hbar$ is the gyromagnetic ratio for electrons, and $M_V=\mu_0 M_s \Omega$. Using spherical coordinate system, with constant magnitude of magnetization, we get the following coupled equations for the dynamics of $\theta(t)$ and $\phi(t)$:
\begin{multline}
\left(1+\alpha^2 \right) \frac{d\theta(t)}{dt} = \frac{|\gamma|}{M_V} \lbrack \{- s \left(c_s(V) + \alpha\,b_s(V) \right) \\ + B_{0e}(t)\} 
sin\theta(t) - 2\alpha B(t) sin\theta (t)cos\theta (t) \rbrack\label{eq:theta_dynamics}
\end{multline}
\begin{multline}
\left(1+\alpha^2 \right) \frac{d\phi(t)}{dt} = \frac{|\gamma|}{M_V} \lbrack \{ s \left(b_s(V)-\alpha\,c_s(V)\right) + \alpha B_{0e}(t)\} \\+ 2 B(t) cos\theta(t) \rbrack \qquad (sin\theta \neq 0).
\label{eq:phi_dynamics}
\end{multline}

Note that when the magnetization vector is aligned 
 along the easy axis (i.e. $\theta = 0^{\circ}, 180^{\circ}$), the torque due to shape anisotropy, 
$\mathbf{T_E}(t)$ and the torque due to spin-transfer-torque, 
$\mathbf{T_{STT}}(t)$ both vanish (see Equations~\eqref{eq:torque_gradient} and~\eqref{eq:torque_STT}), 
which makes  $d\theta(t)/dt$ as well as $d\phi(t)/dt$ equal to zero. Hence the two mutually anti-parallel
orientations along the easy axis
become ``stable''. If the magnetization is in either of these states, no amount of switching current $I$ can 
budge it (because the spin-transfer-torque vanishes), which is why these two orientations are also 
referred to as ``stagnation points''. Fortunately,
thermal fluctuations can dislodge the magnetization from a stagnation point and enable switching.
We will now show that  there can be a third set of values $(\theta_3, \phi_3)$ 
for $\theta(t)$ and $\phi(t)$ 
for which  both $d\theta(t)/dt$ and $d\phi(t)/dt$ will vanish. 

\begin{figure*}
\centering
\subfigure[]{\label{fig:delay_5ns}\includegraphics[width=2in]{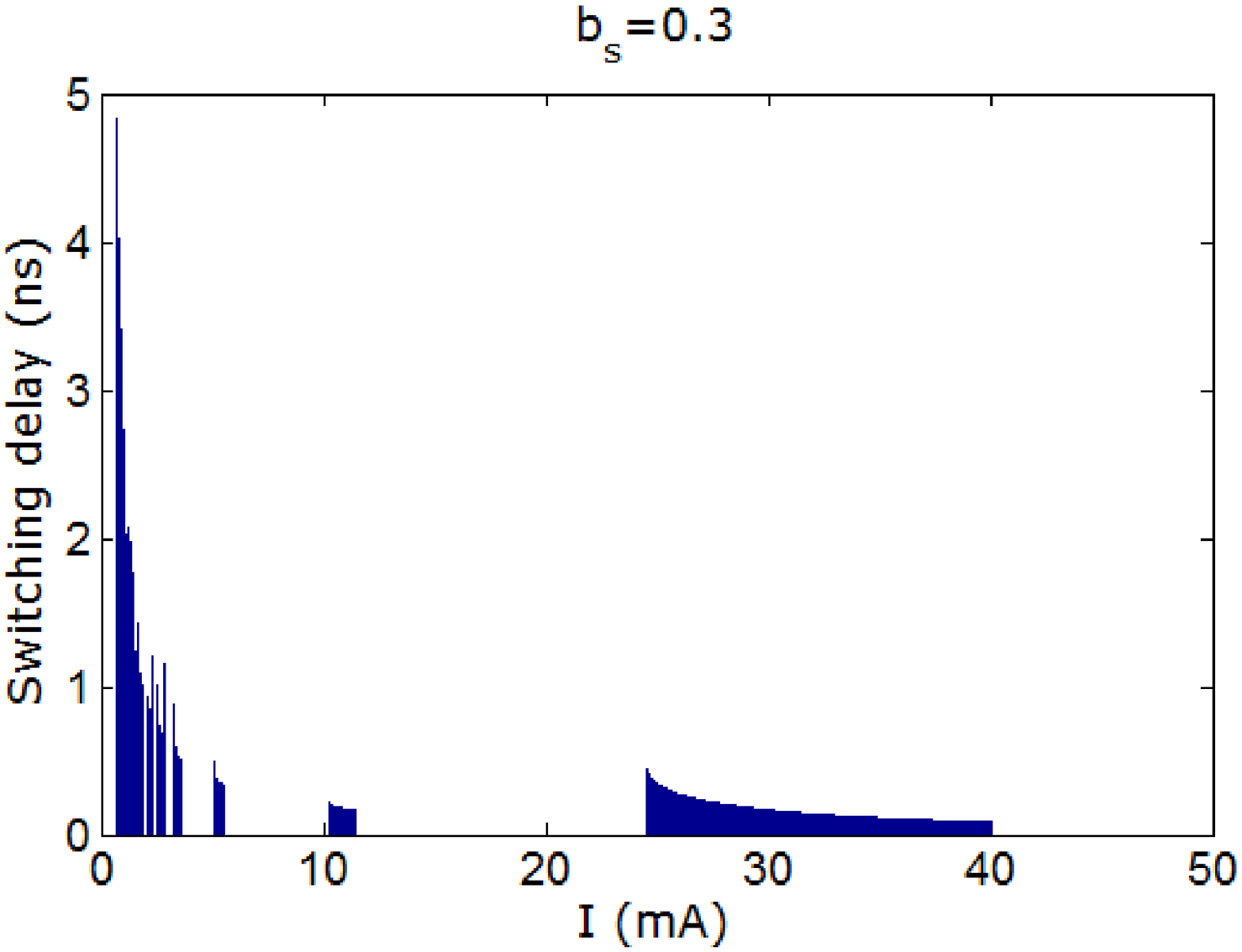}}
\subfigure[]{\label{fig:delay_5ns_b0d05}\includegraphics[width=2in]{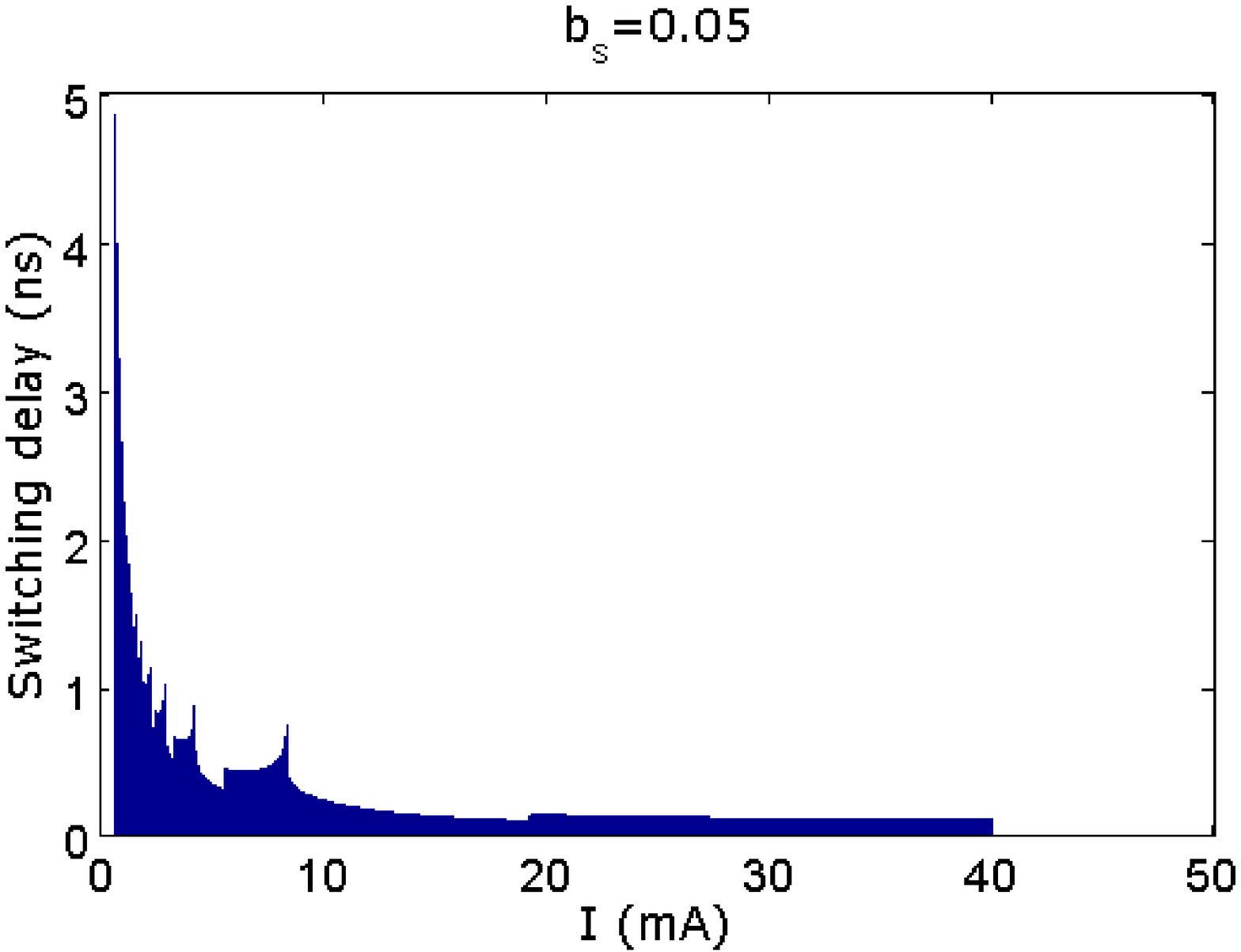}}
\subfigure[]{\label{fig:delay_b0_5ns}\includegraphics[width=2in]{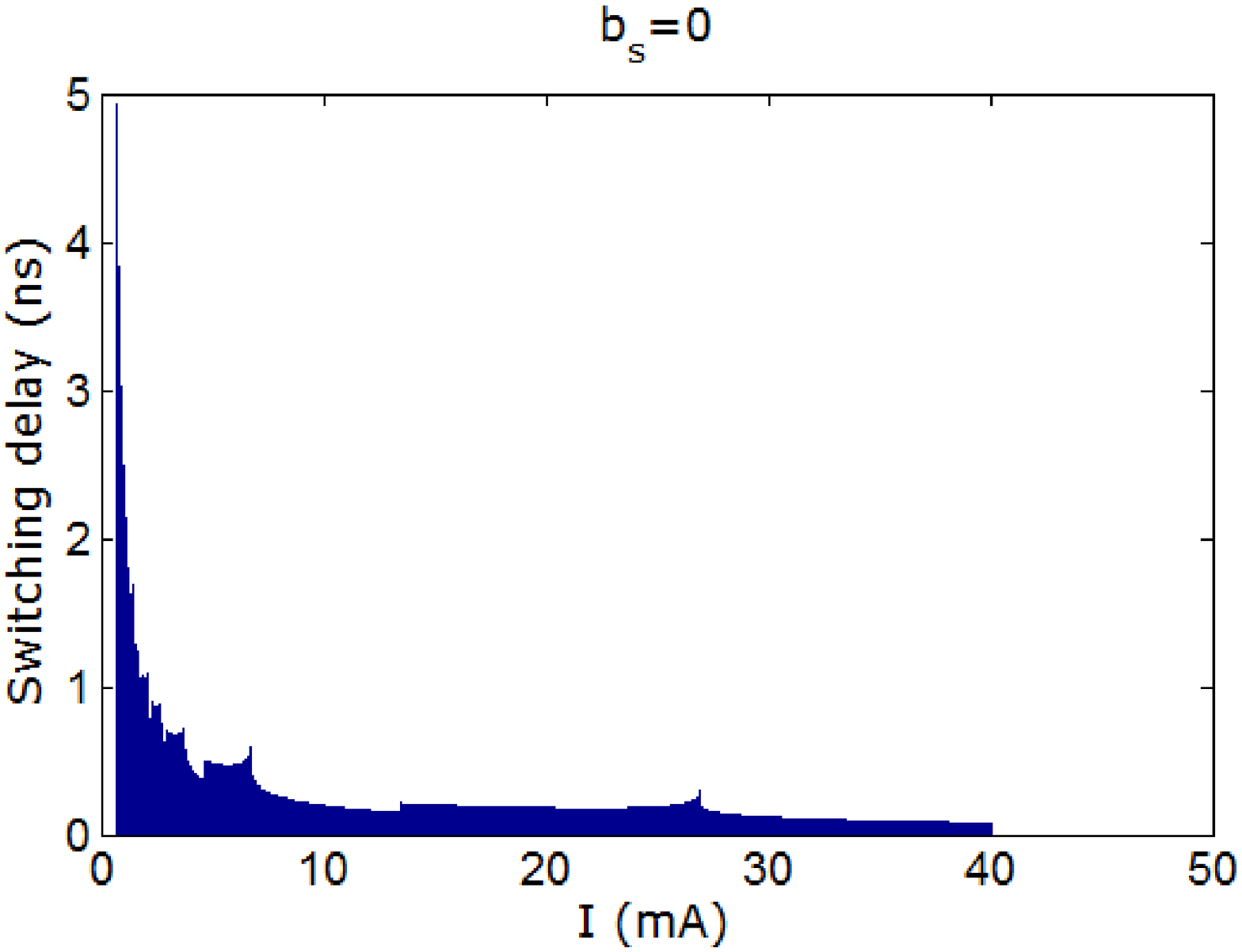}}
\caption{\label{fig:delay_5ns_different_bs} Switching delays for a range of switching current 
700 $\mu$A -- 40 mA producing spin-transfer-torque  for $c_s(V)=-1$ and different values of $b_s(V)$. The switching current is varied in steps of 10 $\mu$A.
(a) $b_s(V)=0.3$.  No switching occurs
 for the following switching current ranges: 2-2.05 mA, 2.33-2.49 mA, 2.83--3.26 mA, 3.69--5.09 mA, 
 5.6--10.24 mA, 11.41--24.51 mA. The switching failure is either due to the metastable state or the magnetization
vector ending up in a state where it stops rotating and begins to oscillate around a mean orientation. The 
latter is a different state not discussed here.
(b) $b_s(V)=0.05$. Switching succeeds for the entire range of switching current. Metastable state appears for $b_s(V) > 0.05$.
(c) $b_s(V)=0$. Switching succeeds for the entire range of switching current.
} 
\end{figure*}

We determine the values of $(\theta_3, \phi_3)$ as follows. 
From Equations~\eqref{eq:theta_dynamics} and~\eqref{eq:phi_dynamics}, 
by making both $d\theta(t)/dt$ and $d\phi(t)/dt$ equal to zero, we get
\begin{align}
2\alpha B(\phi_3) cos\theta_3 &= B_{0e}(\phi_3) - s \, c_s(V) - \alpha s \, b_s(V) \label{eq:theta_dynamics_3} \\
2 B(\phi_3) cos\theta_3 &= - \alpha B_{0e}(\phi_3) + \alpha s \, c_s(V) - s \, b_s(V). \label{eq:phi_dynamics_3}
\end{align}
\noindent
From the above two equations, we get $B_{0e}(\phi_3) = s c_s(V)$. If we put $B_{0e}(\phi_3) = s c_s(V)$ in  Equation~\eqref{eq:theta_dynamics_3} or in Equation~\eqref{eq:phi_dynamics_3}, we get $2 B(\phi_3) cos\theta_3 = - s b_s(V)$. Accordingly, we can determine the values of $(\theta_3, \phi_3)$ as 
\begin{align}
\phi_3 &= \cfrac{1}{2}\, sin^{-1} \left(\cfrac{(\hbar/2e)\eta I c_s(V)}{(\mu_0/2) \, 
M_s^2 \Omega (N_{d-xx}-N_{d-yy}) } \right) \label{eq:phi_3} \\
\theta_3 &= cos^{-1} \left(-\cfrac{(\hbar/2e)\eta I b_s(V)\left [ \mu_0 M_s^2 \Omega \right ]^{-1}}{N_{d-xx}cos^2 \phi_3 
+ N_{d-yy}sin^2 \phi_3 - N_{d-zz} }  \right). \label{eq:theta_3}
\end{align}
\noindent
Note that $\phi_3$ depends on $c_s(V)$ 
while $\theta_3$ depends on both $c_s(V)$ and $b_s(V)$. Neither depends on the Gilbert damping factor $\alpha$.

In order to understand the physical origin of the state $(\theta_3, \phi_3)$, consider the fact that the total torque $\mathbf{T_{eff}}(t) = \mathbf{M}(t) \times \mathbf{H_{eff}}(t)$ can be deduced from Equations 
(\ref{eq:torque_gradient}) and (\ref{eq:torque_STT}) as
\begin{eqnarray}
\mathbf{T_{eff}}(t) & = &  \left \lbrace - 2 B(t) cos \theta(t) - s b_s (V) \right \rbrace sin \theta(t) \, \mathbf{\hat{e}_\phi}
\nonumber \\ 
&& + \left \lbrace- B_{0e}(t) + s c_s(V) \right \rbrace sin \theta(t) \, \mathbf{\hat{e}_\theta}.
\end{eqnarray}
We immediately see that
$\mathbf{T_{eff}}(t)$ vanishes when $\theta(t) = \theta_3$ and $\phi(t) = \phi_3$. Hence there is no torque 
acting on the magnetization vector if it reaches the state $\theta(t) = \theta_3$ and $\phi(t) = \phi_3$
at the same instant of time $t$. Thereafter, it cannot rotate any further since the torque has vanished. Unlike in the case of the 
other two stable states where both shape-anisotropy torque and spin-transfer-torque {\it individually}
vanish, here neither vanishes, but they are equal and opposite so that they cancel to 
make the net torque zero. If the magnetization ends up in this
orientation, then it will be stuck and not rotate further unless we change the switching current $I$
to change the spin-transfer-torque. 
Since changing $I$ can dislodge the magnetization from this state, it is 
{\it not} a stagnation point unlike $\theta = 0^{\circ}, 180^{\circ}$. 
 Hence, we call it a ``metastable'' state.

If $b_s(V)=0$, then $\theta_3=90^\circ$, which means that 
the magnetization will be stuck somewhere in the x-y plane perpendicular to the easy axis, if
it lands in the metastable state. This plane is 
defined by the in-plane and out-of-plane hard axes. Note that when $c_s(V)$ is negative, $\phi_3$ is in the range
$-90^{\circ} 
\leq \phi_3 \leq 0^{\circ}$, but when $c_s(V)$ is positive, $0^{\circ} 
\leq \phi_3 \leq 90^{\circ}$. The quantity $b_s(V)$ cannot be negative~\cite{RefWorks:112}. Note also that when $I$ = 0 so that there 
is no spin-transfer-torque, $\theta_3$ = 90$^{\circ}$ and $\phi_3$ can be any of the following values: $\lbrace 0^{\circ},90^{\circ},180^{\circ},270^{\circ} \rbrace$.
Consequently, the magnetization vector is either along the in-plane hard axis (y-axis) or the out-of-plane hard axis (x-axis).
Since these are obviously metastable states in an unperturbed shape-anisotropic nanomagnet, we call the state 
$\left ( \theta_3, \phi_3 \right )$ a
``metastable'' state.


For numerical simulations, we consider a nanomagnet of elliptical cross-section made of CoFeB alloy 
which has saturation magnetization $M_s=8\times10^5$ A/m~\cite{RefWorks:422} and a 
Gilbert damping factor $\alpha$ = 0.01. We assume the lengths of major axis ($a$), minor axis ($b$), 
and thickness ($l$) to be 150 nm, 100 nm, and 2 nm, respectively. These dimensions ($a$, $b$, and $l$) 
ensure that the nanomagnet will consist of a single ferromagnetic domain~\cite{RefWorks:402,RefWorks:133}. The combination of the parameters $a$, $b$, $l$, and $M_s$ makes 
the in-plane shape anisotropy energy barrier height $\sim$32 kT at room temperature. With the dimensions ($a$, $b$, and $l$) chosen, the demagnetization factors ($N_{d-xx}$,$N_{d-yy}$,$N_{d-zz}$) turn out to be (0.947,0.034,0.019)~\cite{RefWorks:402}. The spin polarization 
of the switching current is always assumed to be 80\%.
 
We assume that the magnetization is initially along the +z-axis, which is a stagnation point. 
Hence, at 0 K, no switching can occur. However, at a finite temperature, thermal fluctuations will 
dislodge the 
magnetization from the stagnation point and enable switching. At room temperature, the thermal fluctuations will deflect the magnetization vector by $\sim$4.5$^{\circ}$ from the easy axis when averaged over time~\cite{roy11_4_2}, so that we will assume the initial value of the polar angle to be
 $\theta_{init}$ = $4.5^\circ$. We choose the initial azimuthal angle $\phi_{init}$ as 
 $+90^\circ$ because it is the most likely value in the absence of spin transfer torque. 
 Similar assumptions are made by others~\cite{RefWorks:403}. We then solve Equations 
 (\ref{eq:theta_dynamics})
 and (\ref{eq:phi_dynamics}) simultaneously to find $\theta(t)$ and $\phi(t)$ as a function of time.
 Once $\theta(t)$ reaches $175.5^\circ$, regardless of the value of $\phi(t)$, we consider the 
 switching to have completed. The time taken for this to happen is the switching delay. 

Fig.~\ref{fig:delay_5ns_different_bs} shows the switching delays 
versus switching current for different values of $b_s(V)$. The switching delay is `infinity' in some current ranges 
when $b_s(V) = 0.03$ because switching failed [see Fig.~\ref{fig:delay_5ns}]. However,
beyond the current of 24.51 mA,  switching always occurs within a finite time, meaning that the 
magnetization never ends up in the metastable state. Simulation results show that if the value of $b_s$ is small enough ($\leq 0.05$), the metastable state does not show up [see Figs.~\ref{fig:delay_5ns_b0d05} and~\ref{fig:delay_b0_5ns}].

The important question is why switching fails only for certain ranges of the current $I$, 
i.e. why does the magnetization vector land in the metastable state for certain values 
of $I$ and not others?
The answer is that starting from some initial condition $\left ( \theta_{init}, \phi_{init} \right )$, the
angles
$\theta(t)$ and $\phi(t)$ must reach the values $\theta_3$ and $\phi_3$ at the {\it same} instant of time $t$. 
This may not happen for 
any arbitrary $I$. Hence, only certain ranges of $I$ will spawn the metastable state. It is also clear from
Equation (\ref{eq:phi_3}) that 
above a certain value of $I$, there will be no real solution for $\phi_3$ since the argument of the 
arcsin function will exceed unity. This value will be $I_{threshold} = 
\left \lbrack e \mu_0 M_s^2 \Omega \left ( N_{d-xx}-N_{d-yy} \right ) \right \rbrack / \left \lbrack \hbar \eta c_s(V) \right \rbrack$.
By maintaining the magnitude of the switching current above $I_{threshold}$, we can ensure that the 
magnetization vector will never get stuck at the metastable state. For the nanomagnet considered, $I_{threshold}$
= 32.7 mA, but switching becomes feasible at even lower current of 24.52 mA since in the range 
[24.52 mA, 32.7 mA], the coupled $\theta$ and $\phi$-dynamics expressed by Equations (\ref{eq:theta_dynamics})
and (\ref{eq:phi_dynamics}) do not allow $\theta(t)$ and $\phi(t)$ to reach $\theta_3$ and $\phi_3$ 
{\it simultaneously} starting from $\left ( \theta_{init}, \phi_{init} \right )$.

\begin{figure*}
\centering
\subfigure[]{\label{fig:theta_dynamics_stt_24d51mA_thermal}\includegraphics[width=2in]
{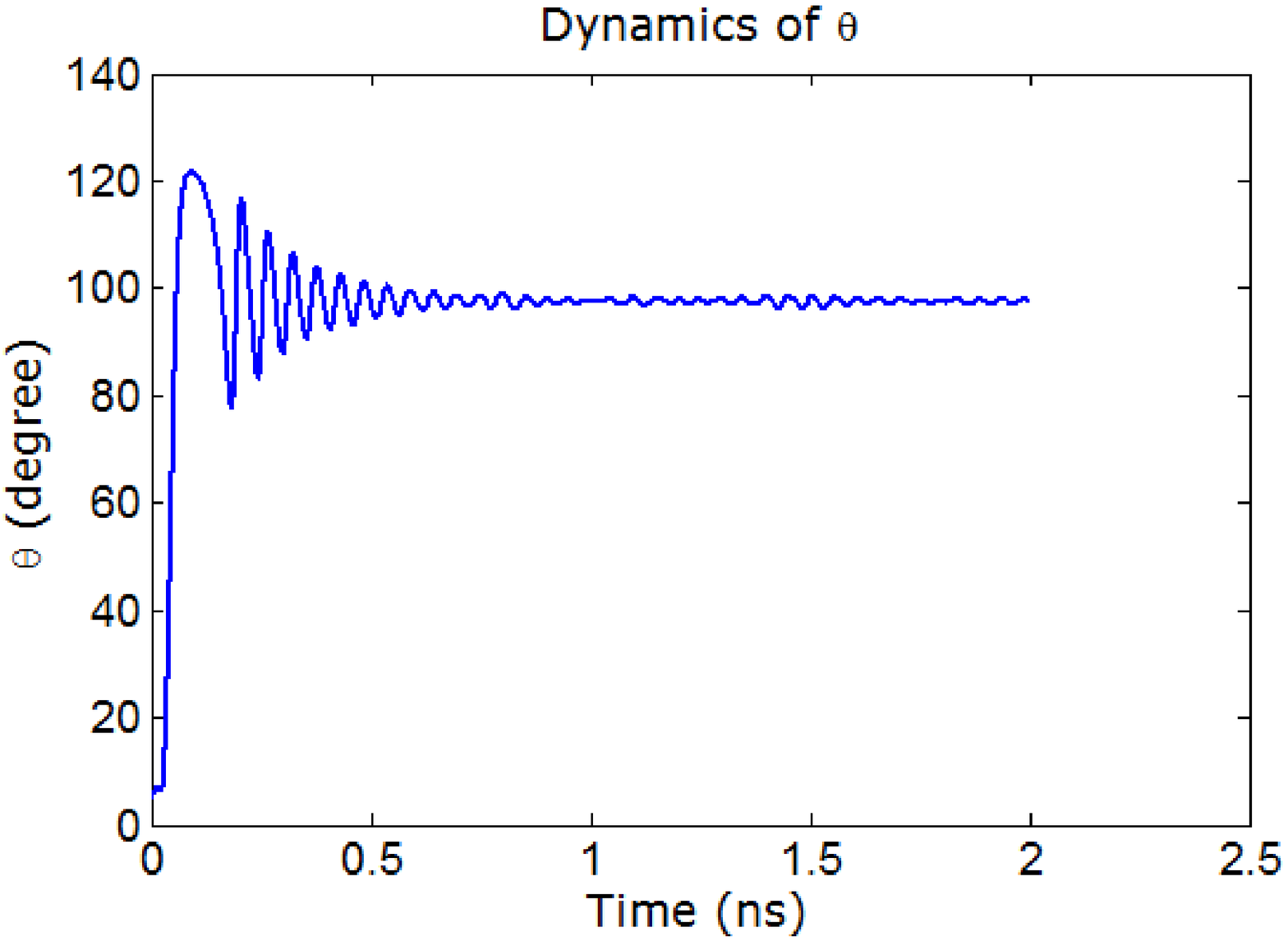}}
\subfigure[]{\label{fig:phi_dynamics_stt_24d51mA_thermal}\includegraphics[width=2in]
{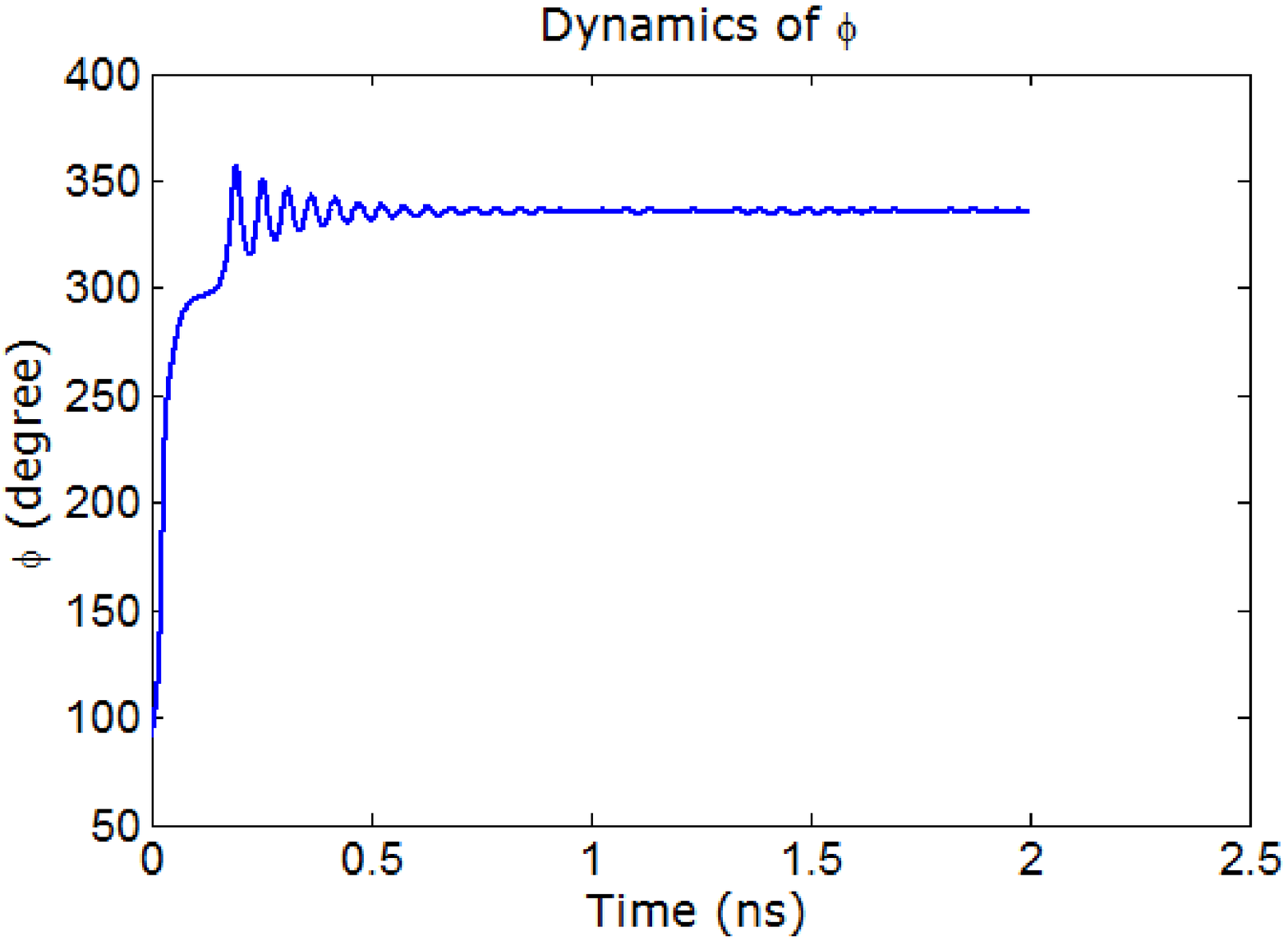}}
\subfigure[]{\label{fig:dynamics_stt_24d51mA_thermal}\includegraphics[width=2in]
{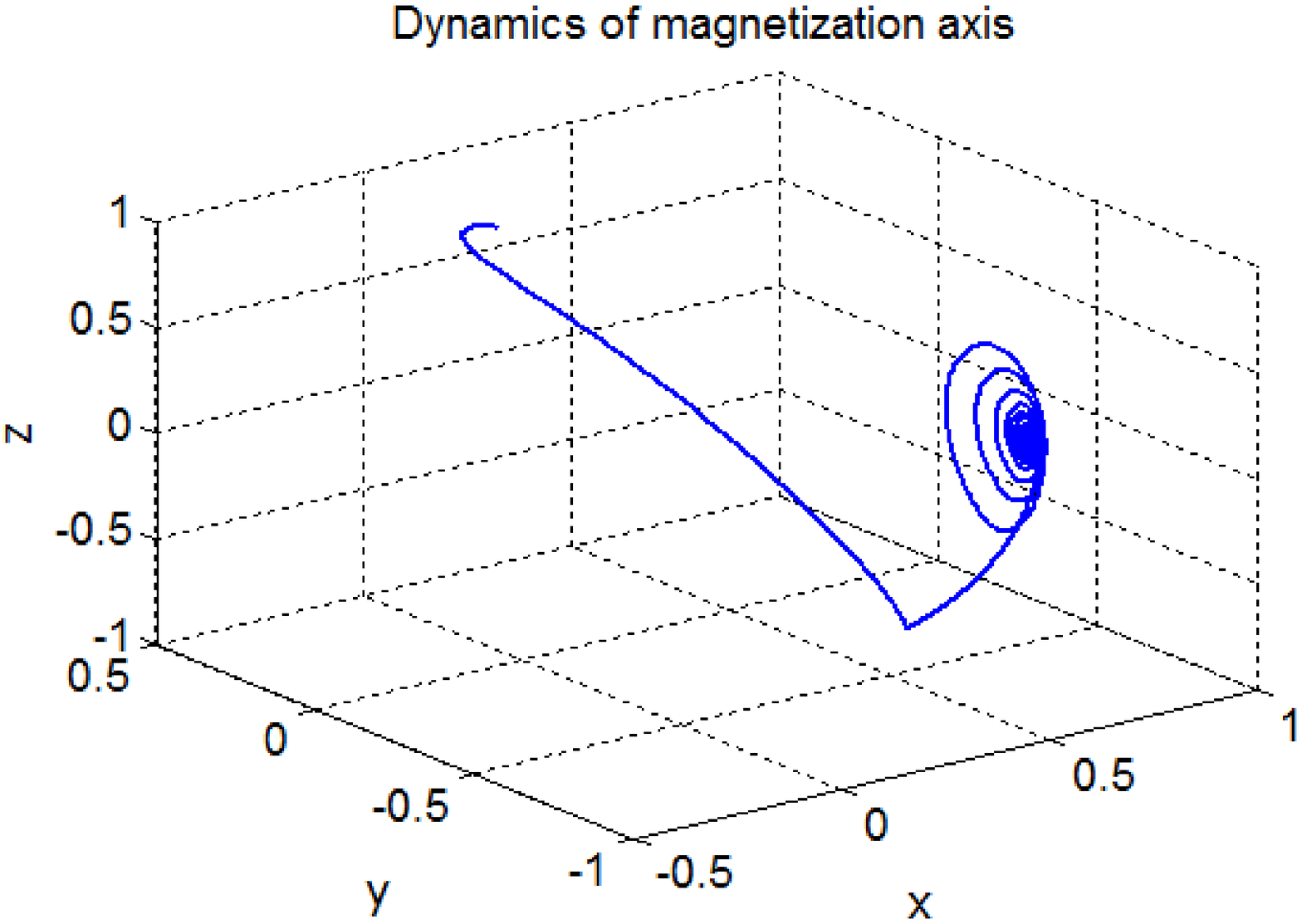}}
\caption{\label{fig:dynamics_stt_24d51mA_thermal_all} Room-temperature (300 K) magnetization dynamics when the switching current is 24.51 mA ($c_s(V)=-1$, $b_s(V)=0.3$). (a) 
Dynamics of $\theta(t)$. (b) Dynamics of  $\phi(t)$. (c) The trajectory traced out by the tip of the magnetization vector in three-dimensional space.
The magnetization gets stuck at a metastable state with $\theta_3$ = 97.58$^\circ$ and 
$\phi_3$ = 335.87$^\circ$. This plot was obtained by the solution of the stochastic Landau-Lifshitz-Gilbert equation in
the presence of a random thermal torque to simulate the effect of thermal fluctuations~\cite{roy11_4_2}. This is one specific run from 10,000 simulations performed in the presence of thermal fluctuations that shows that the latter cannot untrap the magnetization from this state at room temperature. This happens for all the 10,000 simulations if thermal fluctuations are brought into play after the metastable state is reached. This shows that the state is stable against room-temperature thermal perturbations. }
\end{figure*}

\begin{figure}
\centering
\includegraphics[width=2in]{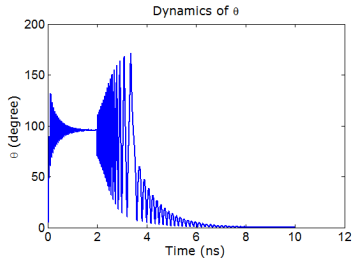}
\caption{\label{fig:theta_dynamics_stt_20mA_Ioff_2ns} Time evolution of $\theta(t)$ when the switching current is 
20 mA ($c_s(V) = -1$, $b_s(V) = 0.3$). Thermal fluctuations were ignored but we assumed $\theta_{init} = 4.5^{\circ}$ and $\phi_{init} = 90^{\circ}$, respectively. The switching current is turned off at 2 ns after the magnetization vector gets stuck at the
 metastable state with $\theta_3$ = 95.74$^\circ$ and 
$\phi_3$ = 341.25$^\circ$. The dynamics shows that the magnetization vector relaxes to the easy axis because of shape anisotropy, but the 
final orientation is the {\it wrong} orientation  along
the +z-axis rather than the desired final orientation along the -z-axis. Therefore, switching fails.} 
\end{figure}

Another important question is whether thermal fluctuations can untrap the magnetization from this state.
To probe this, we solved the stochastic LLG equation~\cite{roy11_4_2} in the presence of a random thermal torque. Fig.~\ref{fig:dynamics_stt_24d51mA_thermal_all} shows the magnetization dynamics for a switching current of 24.51 mA at room temperature (300 K). 
We observe that the magnetization gets stuck at a metastable state (with $\theta_3$ = 97.58$^\circ$ and 
$\phi_3$ = 335.87$^\circ$) $\sim$50\% of the time, which means that roughly one-half of the 
switching trajectories intersect the metastable state and terminate there. The values of
$\theta_3$, $\phi_3$ are also the angles predicted by
Equations~\eqref{eq:phi_3} and~\eqref{eq:theta_3}, thereby confirming that the metastable 
state indeed has the origin described here. Increasing the temperature to 400 K helps by decreasing the probability that a switching trajectory will intersect the metastable state~\cite{supplx}.
What is important however is that {\it if}  the 
magnetization vector gets stuck in the metastable state and the current remains on,
then no amount of thermal fluctuations can dislodge it. In other words, this state is {\it stable} against 
thermal perturbations.

Unfortunately, an analytical stability analysis is precluded by the complex coupled $\theta$-$\phi$ dynamics. Therefore, we performed numerical stability analysis while spanning the parameter space
as exhaustively as possible. In all cases, we found the state to be stable against thermal agitations.

We also notice that oscillations precede settling into the
metastable state. This is due to coupled $\theta$- and $\phi$-dynamics governing the rotation of the 
magnetization vector, which causes some ringing.
The metastable state appears in the switching current range 11.41 mA to 24.51 mA since 
within this range,
$\theta(t)$ and $\phi(t)$ can reach $\theta_3$ and $\phi_3$ simultaneously starting with the initial 
conditions $\left ( \theta_{init}, \phi_{init} \right )$. If the initial conditions are changed,
the range can change as well~\cite{supplx}.

Finally, one issue that merits discussion is what happens if the spin polarized current is turned off after the 
magnetization gets stuck. In that case, the torque due to shape anisotropy will take over and drive the 
magnetization to the easy axis. One expects that if $\theta_3 < 90^{\circ}$, then switching will fail since the 
nearer easy axis is the undesired orientation, whereas if $\theta_3 > 90^{\circ}$,
then switching should succeed because the nearer easy axis is the desired orientation. Equation (\ref{eq:theta_3}) dictates that $\theta_3 > 90^{\circ}$ since $b_s(V)$ is always positive. Unfortunately, these simple
expectations are belied by the complex dynamics of magnetization. The out-of-plane excursion of the magnetization 
vector causes an additional torque that depends on $\theta_3$, $\phi_3$. The torque can oppose the torque due to shape anisotropy. As a result,
even when $\theta_3 > 90^{\circ}$, switching can fail since the magnetization reaches the wrong orientation along the 
easy axis (see Fig.~\ref{fig:theta_dynamics_stt_20mA_Ioff_2ns}).

In conclusion, we have shown the existence of a new metastable magnetization state in a single-domain nanomagnet with
uniaxial shape anisotropy carrying a spin-polarized current. If the magnetization gets trapped in this state, switching will
fail with non-zero probability even in the presence of thermal fluctuations. This has vital implications for STT-RAM technology.


%

\end{document}


\maketitle

In this supplementary material, we provide additional data and plots for magnetization dynamics with corresponding descriptions to elucidate the 
physics and behavior of the metastable state.
It is well known that there is a minimum (threshold) current needed to switch magnetization with spin-transfer-torque that depends on various 
magnet parameters such as shape anisotropy energy barrier, etc.
[Ref. 2 in the main Communication]. The current ranges that we have considered here are above this minimum 
current. The purpose is to show that merely exceeding the minimum  current does {\it not} guarantee switching
because of the existence of the metastable state. Some excess current is needed to guarantee {\it fail-safe}
switching.

In the main Communication, we showed that there is a range of current for which switching can fail for one specific initial condition 
$\theta_{init}=4.5^\circ$ and $\phi_{init} = 90^\circ$. In this supplement, we show that failure can occur for other initial conditions 
as well. Thus, the metastable state is a threat for almost any initial condition. 

Another issue we explore in depth is whether turning off the current once the magnetization gets stuck at the metastable 
will ensure that the shape anisotropy takes over and eventually switches the magnetization to the correct state, 
albeit with a long switching delay. We present simulations to show that this is not the case, although one might 
intuitively think otherwise. Even if the magnetization gets stuck at the metastable state after crossing the 
in-plane hard axis ($\theta_3 > 90^\circ$), turning off the current at that point will not necessarily make it relax towards 
$\theta  = 180^\circ$ and successfully complete switching. The strong coupling between the  $\theta$- and $\phi$-dynamics 
can cause the magnetization to rotate backwards (owing to the strong counter-clockwise torque produced by the out-of-plane 
excursion of the magnetization vector for certain values of $\phi$) and actually end up in the wrong state. These results 
show that the metastable state can be a serious spoiler in STT based switching.

\clearpage
\pagebreak
\begin{figure}
\centering
\includegraphics[width=6in]{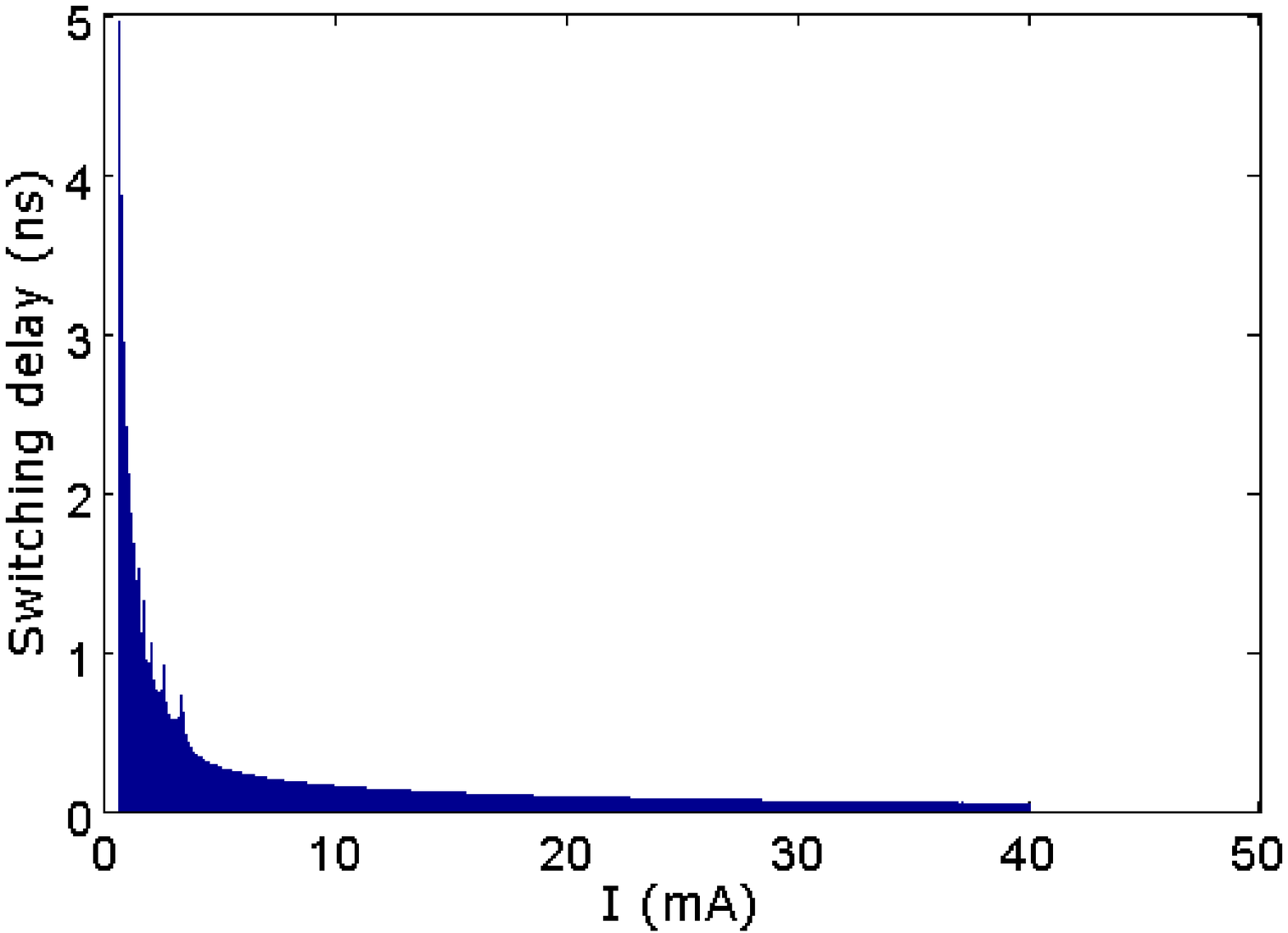}
\caption{\label{fig:delay_b_reversed_5ns} Switching delay versus switching current in the range 700 $\mu$A -- 40 mA. 
Thermal fluctuations were ignored in the magnetization dynamics, but the initial values of $\theta$ and $\phi$ were
assumed to be 4.5$^{\circ}$ and 90$^{\circ}$, respectively.
Any lower switching current will increase the delay to 
more than 5 ns and is hence not considered.
We used the parameters $c_s(V)=+1$, $b_s(V)=0.3$ in this plot. The positive sign of $c_s(V)$ signifies $\theta$ = 180$^\circ$ to 0$^\circ$ switching.
Unlike when $c_s=-1$ ($\theta$ = 0$^\circ$ to 180$^\circ$ switching), the switching delay always remains finite showing that the magnetization 
vector never gets stuck at the metastable state within this range of switching current
for anti-parallel to parallel switching.  Changing $\phi_{init}$ did not make any difference. 
This did not happen because there is no metastable state in this case, but because  $\theta(t)$ and $\phi(t)$ cannot reach the values $\theta_3$ and $\phi_3$ simultaneously within this current range.
Thus, even though the metastable state exists, it does not hinder switching since the magnetization vector never visits this state during its sojourn
from one stable state along the easy axis to the other. In other words, the switching trajectory
does not intersect the metastable state.} 
\end{figure}

\clearpage
\pagebreak
\begin{figure}
\centering
\includegraphics[width=6in]{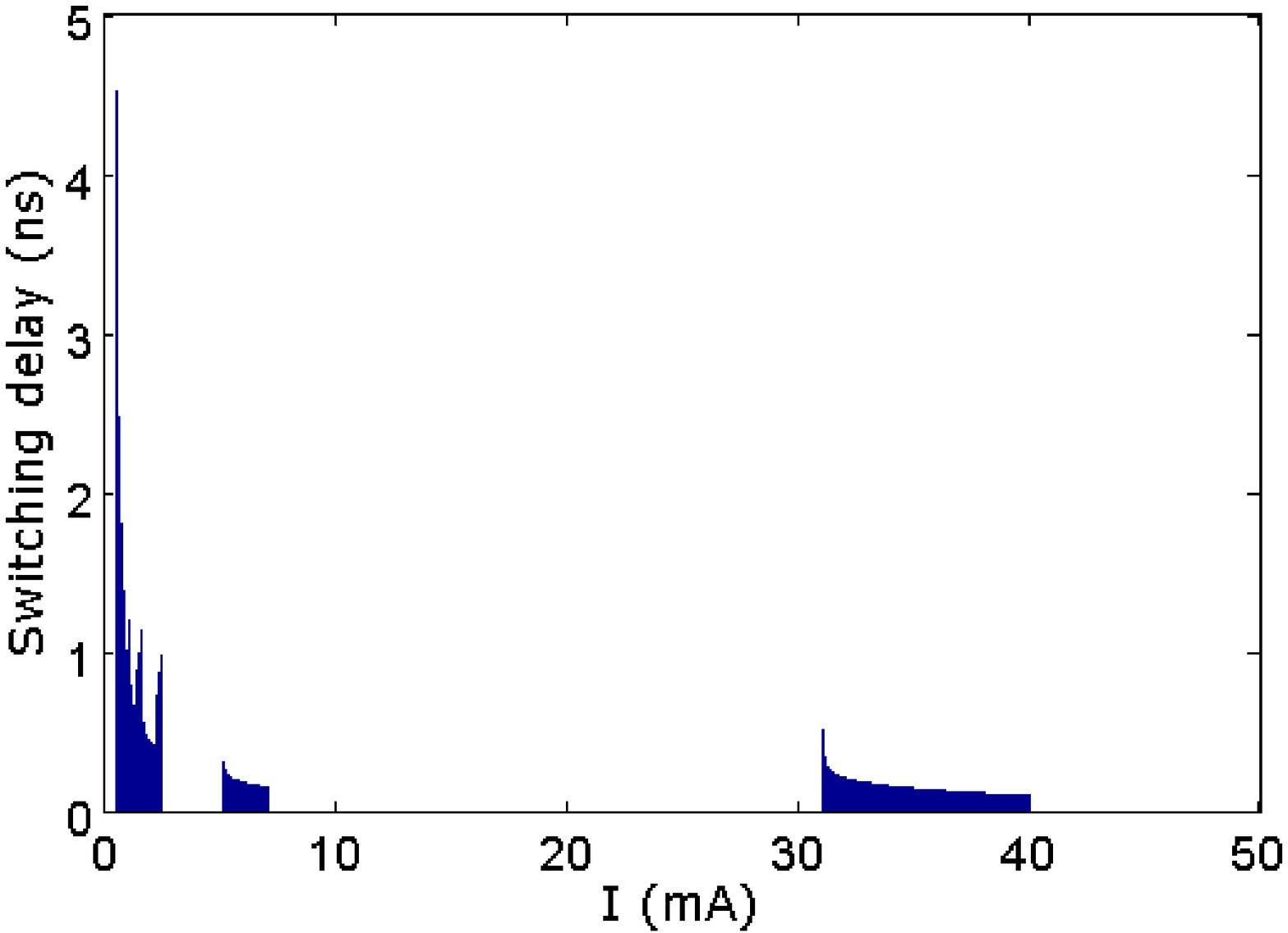}
\caption{\label{fig:delay_phi_0deg_5ns} Switching delay versus switching 
current in the range 700 $\mu$A -- 40 mA. Thermal fluctuations were ignored in the magnetization dynamics, but the initial values of $\theta$ and $\phi$ were
assumed to be 4.5$^{\circ}$ and 0$^{\circ}$, respectively.
We used the parameters $c_s(V)=-1$ and $b_s(V)=0.3$ in this plot. This corresponds to
parallel to anti-parallel switching.
No switching takes place for the following ranges of switching currents: 
 2.49-5.18 mA, 7.1-31.08 mA.
The switching dynamics was simulated for every 10 $\mu$A of switching current in this interval. 
 Note that the ranges of current where switching fails are sensitive 
to initial conditions since these ranges are different from those in Figure 2 of the main Communication, where the 
initial condition was $\phi_{init}=90^\circ$ and $\theta_{init}=4.5^\circ$.} 
\end{figure}

\clearpage
\pagebreak
\begin{figure}
\centering
\includegraphics[width=6in]{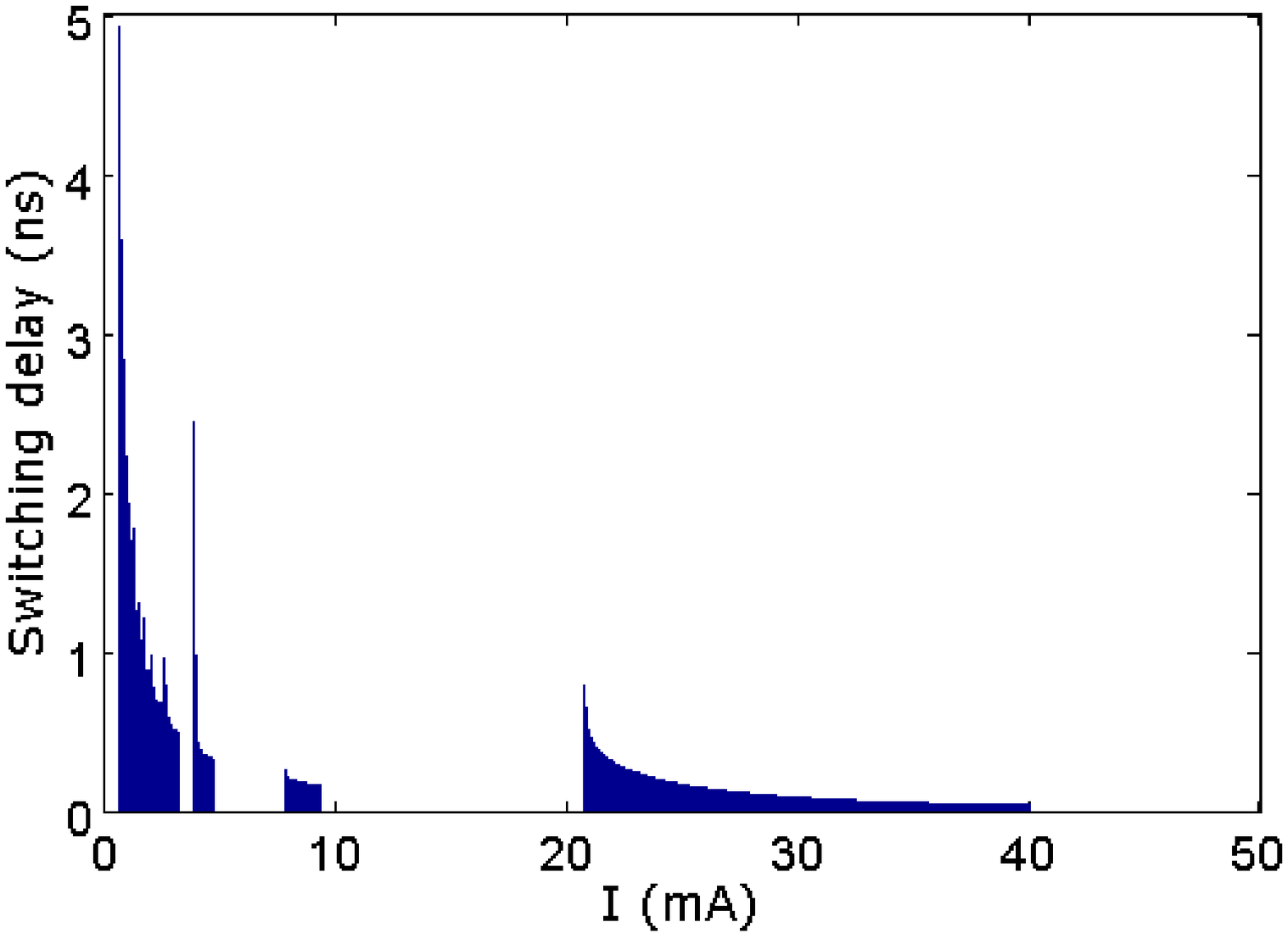}
\caption{\label{fig:delay_theta_7deg_5ns} Switching delay versus switching 
current in the range 770 $\mu$A -- 40 mA. Thermal fluctuations were ignored in the magnetization dynamics, but the initial values of $\theta$ and $\phi$ were assumed to be 7$^{\circ}$ and 90$^{\circ}$, respectively.
We used the parameters $c_s(V)=-1$ and $b_s(V)=0.3$ in this plot. 
No switching takes place for the following ranges of switching currents: 
 2.08 mA, 2.5-2.66 mA, 3.26-3.91 mA, 4.82-7.88 mA, 9.45-20.79 mA. 
The switching dynamics was simulated for every 10 $\mu$A of switching current in this interval. 
 Note once again that the ranges of current where switching fails are sensitive 
to initial conditions since these ranges are different from those in Figure 2 of the main Communication, where the 
initial condition was $\phi_{init}=90^\circ$ and $\theta_{init}=4.5^\circ$.} 
\end{figure}

\clearpage
\pagebreak
\begin{figure}
\centering
\includegraphics[width=6in]{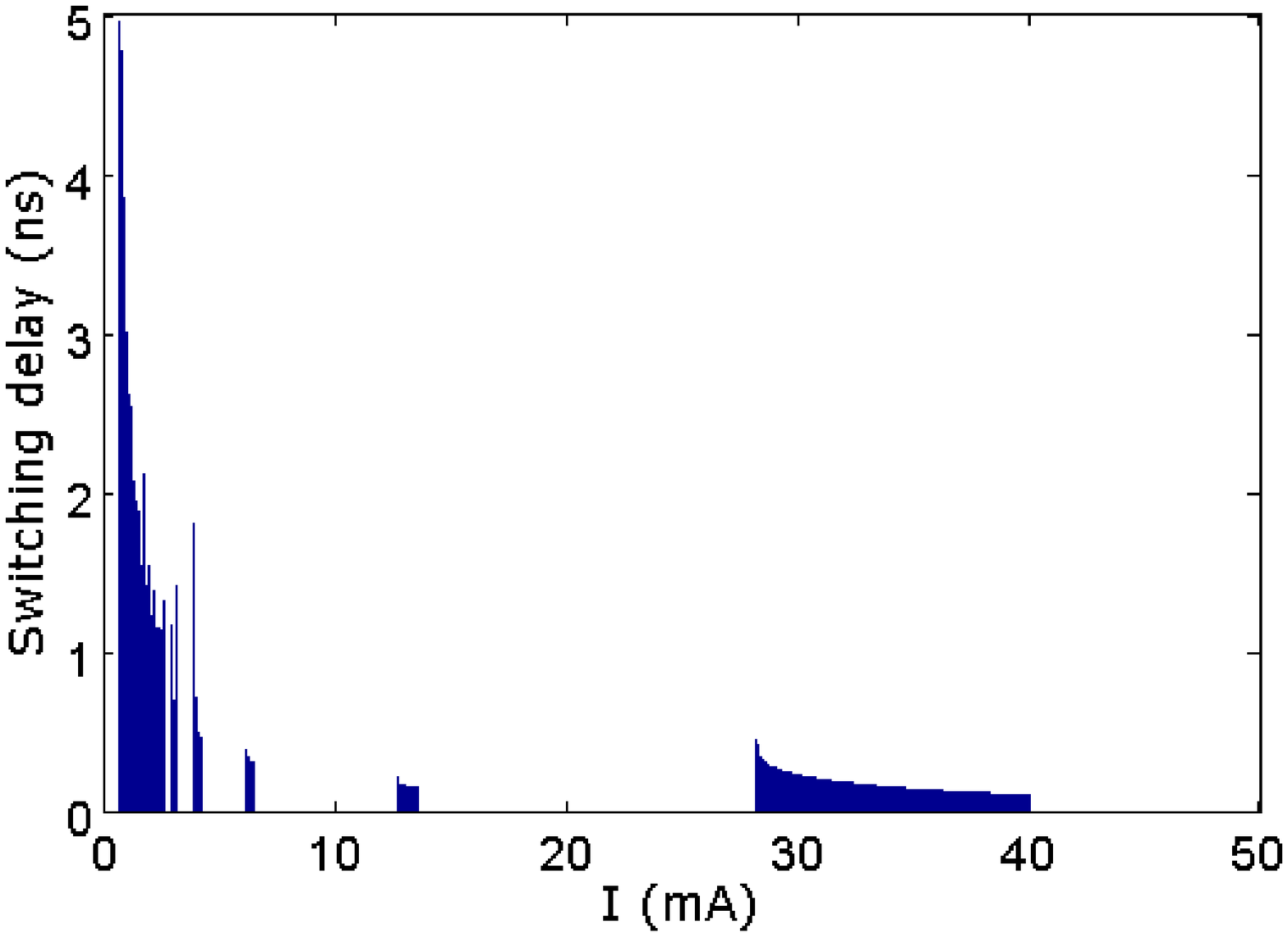}
\caption{\label{fig:delay_theta_3deg_5ns} Switching delay versus switching 
current in the range 770 mA -- 40 mA. Thermal fluctuations were ignored in the magnetization dynamics, but the initial values of $\theta$ and $\phi$ were assumed to be 3$^{\circ}$ and 90$^{\circ}$.
We used the parameters $c_s(V)=-1$ and $b_s(V)=0.3$ in this plot. 
No switching takes place for the following ranges of switching currents: 
 1.97-2.01 mA, 2.24-2.35 mA, 2.63-2.89 mA, 3.22-3.87 mA, 4.21-6.15 mA, 6.49-12.67 mA, 13.62-28.2 mA.
The switching dynamics was simulated for every 10 $\mu$A of switching current in this interval. 
 Once again note that the ranges of current where switching fails are sensitive 
to initial conditions since these ranges are different from those in Figure 2 of the main Communication, where the 
initial condition was $\phi_{init}=90^\circ$ and $\theta_{init}=4.5^\circ$.} 
\end{figure}

\begin{figure}
\centering
\subfigure[]{\label{fig:theta_dynamics_stt_24d51mA_thermal_400K}\includegraphics[width=4.4in]
{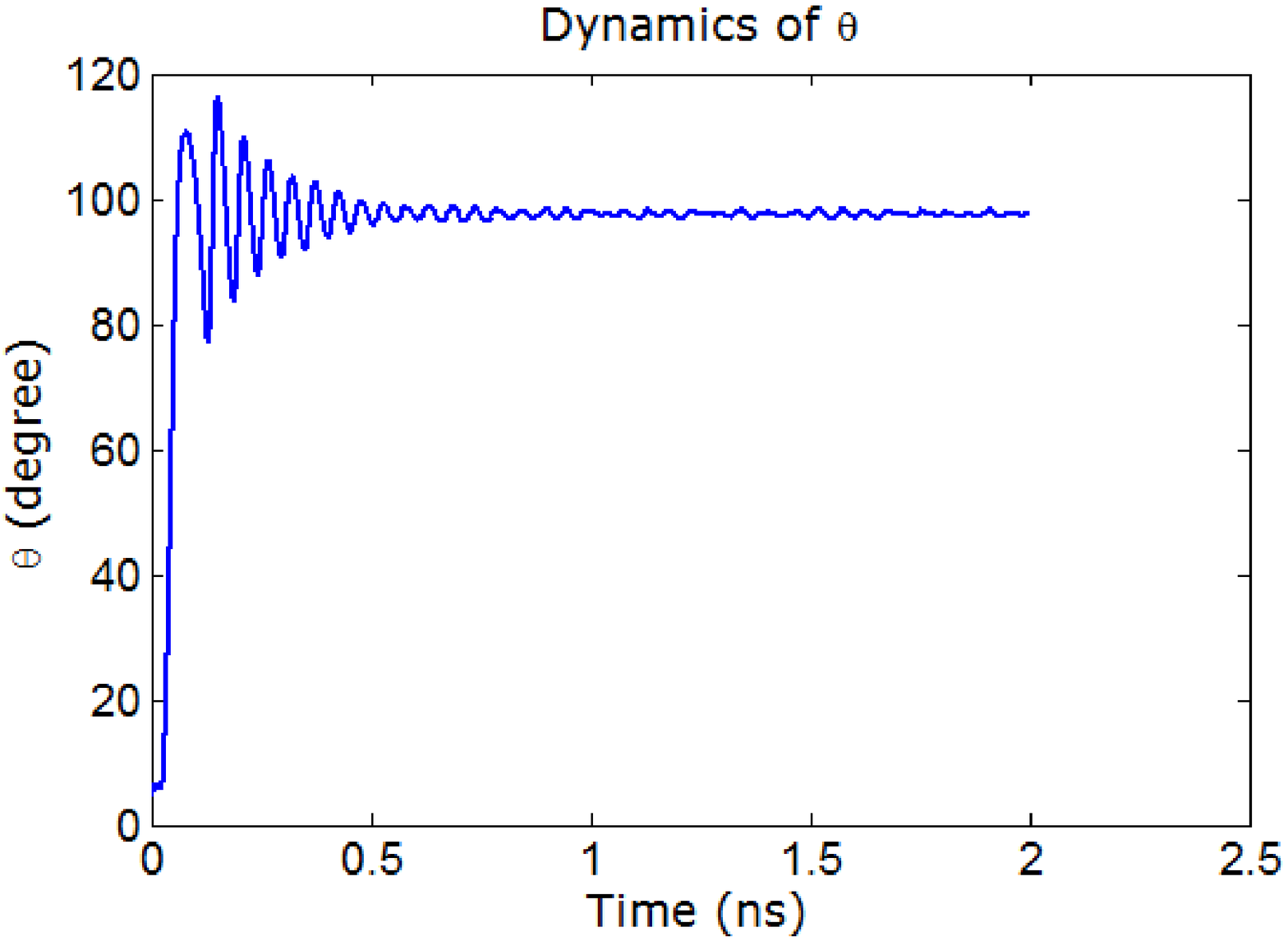}}
\subfigure[]{\label{fig:phi_dynamics_stt_24d51mA_thermal_400K}\includegraphics[width=4.4in]
{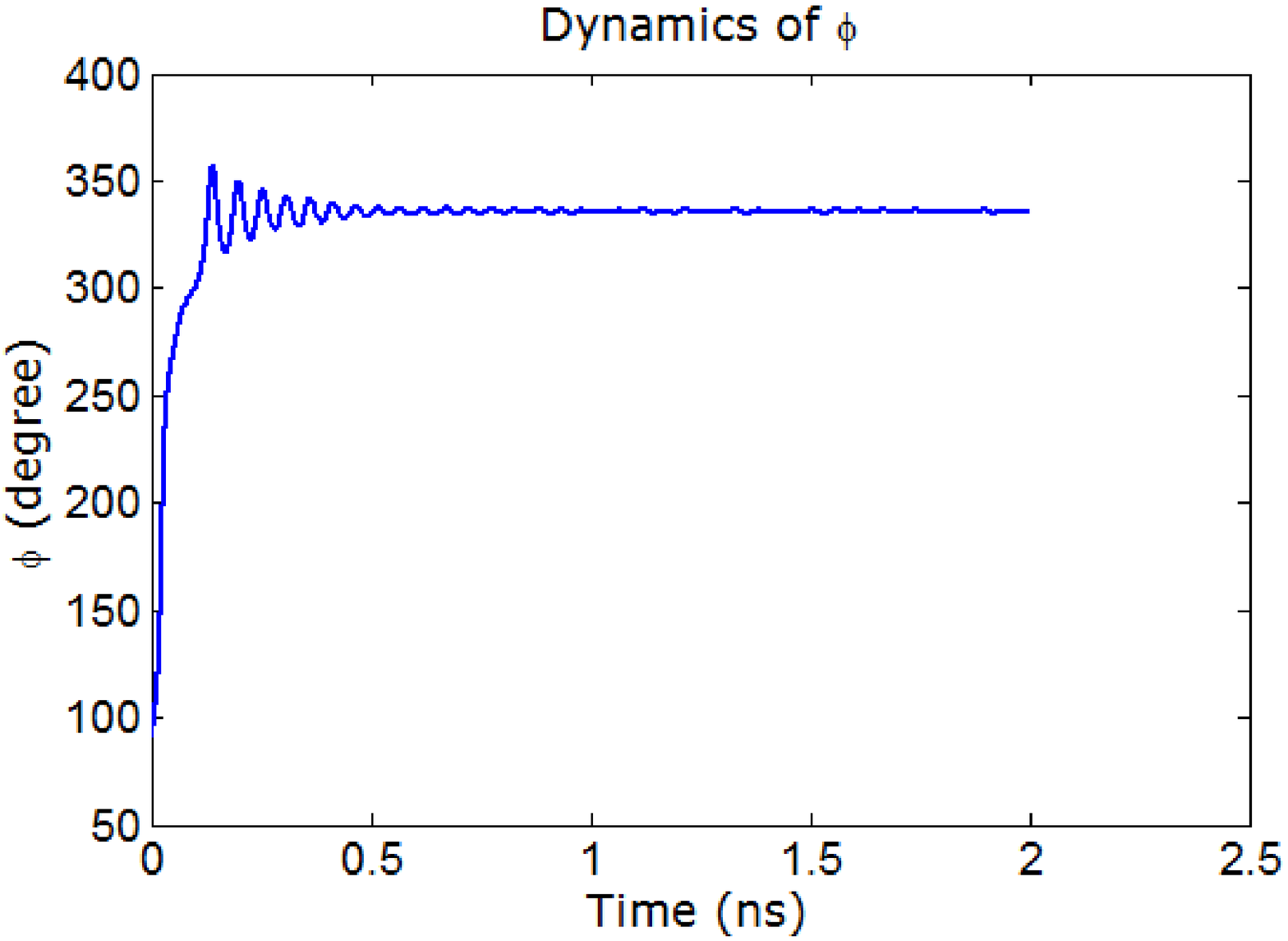}}
\caption{\label{fig:dynamics_stt_24d51mA_thermal_400K} Time evolution of $\theta(t)$ and $\phi(t)$ when the switching current is 
24.51 mA ($c_s(V) = -1$, $b_s(V) = 0.3$) and the temperature is 400 K. 
The magnetization vector gets stuck at the metastable state and cannot be unstuck by thermal fluctuations at 400 K. This plot was obtained by the solution of the stochastic Landau-Lifshitz-Gilbert equation in
the presence of a random thermal torque to simulate the effect of thermal fluctuations. 
Occasional small perturbations around the metastable state show up in both the dynamics of $\theta(t)$ 
and $\phi(t)$ due to thermal fluctuations; however, such small fluctuations are not sufficient 
to dislodge the magnetization vector from the metastable state. Simulation over a long time duration (500 ns)
was carried out to confirm this. This is one specific run from 10,000 simulations in the presence of thermal fluctuations that shows that the latter cannot untrap the magnetization from this state at 400 K. This happens for all the 10,000 simulations if  thermal fluctuations are brought into play {\it after} the metastable state is reached. However, if we consider thermal fluctuations from the very beginning, magnetization does not even reach the metastable state for $\sim$52\% of the simulations and switching occurs successfully, while for the rest $\sim$48\% of the simulations, the magnetization vector gets trapped at the metastable state and thermal fluctuations cannot untrap it thereafter. The failure probability at room temperature (300 K) was $\sim$50\% instead of $\sim$48\%, which shows that increasing the temperature slightly decreases the probability of a switching trajectory intersecting the metastable state,
but in the event such intersection occurs and the magnetization gets stuck, the elevated temperature still cannot untrap it. (a) Dynamics of $\theta(t)$. (b) Dynamics of  $\phi(t)$.}
\end{figure}

\begin{figure}
\centering
\subfigure[]{\label{fig:theta_dynamics_stt_24d51mA_Ioff_2ns}\includegraphics[width=5in]
{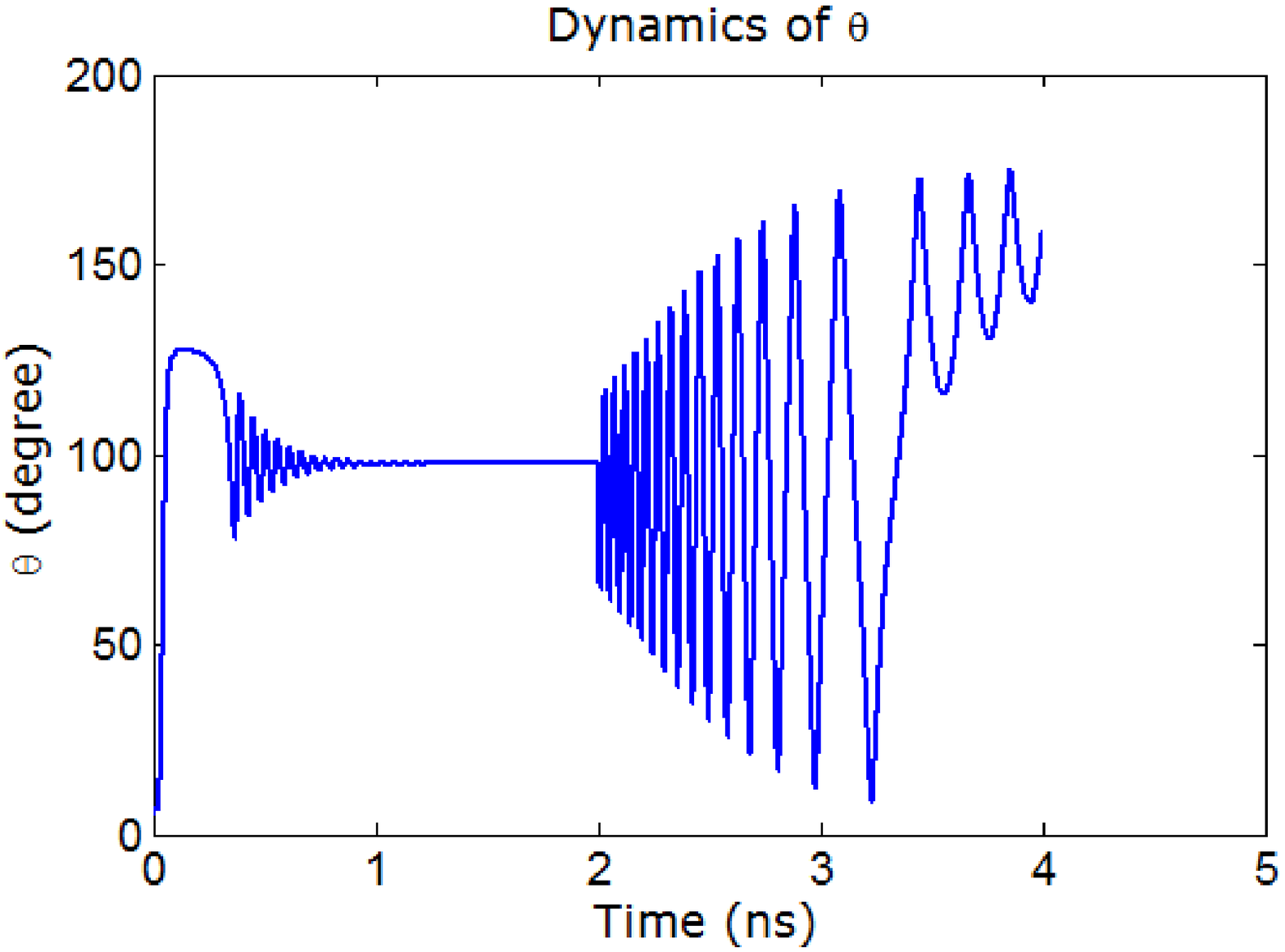}}
\subfigure[]{\label{fig:phi_dynamics_stt_24d51mA_Ioff_2ns}\includegraphics[width=5in]
{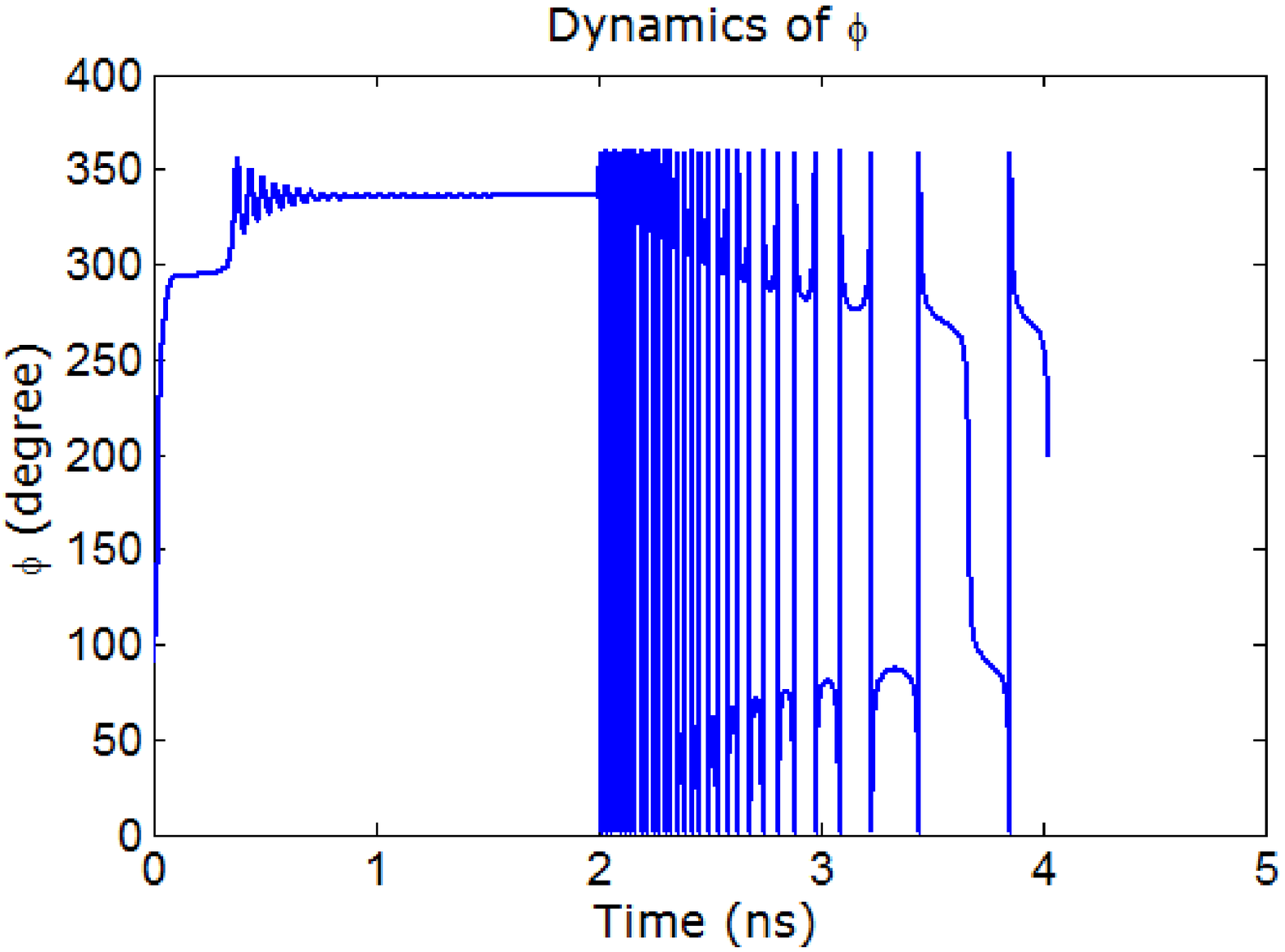}}
\caption{\label{fig:dynamics_stt_24d51mA_Ioff_2ns} Time evolution of $\theta(t)$ and $\phi(t)$ when the switching current is 
24.51 mA ($c_s(V) = -1$, $b_s(V) = 0.3$). Thermal fluctuations were ignored but we assumed $\theta_{init} = 4.5^{\circ}$ and $\phi_{init} = 90^{\circ}$, respectively.
 The switching current is turned off at 2 ns after the magnetization vector gets stuck at the metastable state with $\theta_3$ = 97.58$^\circ$ and 
$\phi_3$ = 335.87$^\circ$. The dynamics shows that magnetization ultimately switches to the correct 
desired orientation along the easy axis ($\theta \approx 180^{\circ}$) because the torque due to shape-anisotropy takes over once the current is switched off; however, the switching takes too long since the shape anisotropy torque is much weaker than the spin-transfer-torque. (a) Dynamics of $\theta(t)$. (b) Dynamics of  $\phi(t)$. This is one case whether turning off the switching current resulted in
successful switching. The other case, discussed in the main Communication, resulted in switching failure. Thus, turning off the switching current once the magnetization gets stuck does not guarantee
successful switching.}
\end{figure}

%